\documentclass{article}
\usepackage{graphicx} 
\usepackage[margin=1in]{geometry}
\usepackage{graphicx}
\usepackage{amsmath}
\usepackage{booktabs}
\usepackage{url}
\usepackage{natbib}

\title{Bots Don't Sit Still: A Longitudinal Study of Bot Behaviour Change, Temporal Drift, and Feature-Structure Evolution}
\author{Ohoud Alzahrani, Russell Beale, Bob Hendley}
\date{School of Computer Science. University of Birmingham, Edgbaston, Birmingham. B15 2TT, UK}
\date{oyzahrani@uqu.edu.sa; R.Beale@bham.ac.uk; R.J.Hendley@bham.ac.uk}

\date{December 2025}

\begin{document}
\maketitle

\begin{abstract}

Social bots are now deeply embedded in online platforms, where they are used for promotion, persuasion, and manipulation. Most bot-detection systems still treat behavioural features as static, implicitly assuming that bots behave in a stationary way over time. This paper asks whether that assumption is justified for promotional bots on Twitter, and examines both how their \emph{individual} behavioural signals change and how the \emph{relationships} between those signals evolve.

Using a longitudinal dataset of 2{,}615 promotional bot accounts and 2.8 million tweets collected between 2009 and 2020 from widely used ground-truth corpora, we first construct yearly time series for ten content-based behavioural meta-features (tweeting, retweeting, replying, URLs, hashtags, duplicated text, sentiment, languages, emojis, and media). We assess stationarity using Augmented Dickey--Fuller and Kwiatkowski--Phillips--Schmidt--Shin tests and estimate linear trends for each feature. All ten meta-features exhibit non-stationary behaviour. Nine show clear upward trends (tweeting, retweeting, replying, URLs, hashtags, duplicated text, sentiment, emojis, media), while language diversity shows a small decline. Stratifying by activation generation (2009--2012, 2013--2016, 2017--2020) and account age (short-, mid-, and long-lived accounts) reveals systematic differences: second-generation bots are the most active and link-heavy, short-lived bots show intense, repetitive activity with heavy hashtag and URL usage, while long-lived bots display lower overall activity but greater language diversity and more varied emotional expression via emojis.

We then complement this with a second study that analyses how behavioural features co-occur within and across three generations of promotional bots. Using an interpretable set of 18 binary features capturing posting actions, topic similarity, URLs, hashtags, sentiment, emojis, and media, we examine 153 pairwise relationships. $\chi^2$ tests show that almost all feature pairs are dependent. Tracking Spearman correlation categories across generations reveals systematic shifts in both strength and polarity: many relationships between key features (e.g.\ multiple hashtags with media, sentiment with URLs) strengthen over time, while others flip from weakly positive to weakly or moderately negative. Later bot generations therefore exhibit more structured and coordinated combinations of behavioural cues.

Taken together, these two studies provide systematic evidence that promotional social bots adapt their behaviour over time at both the level of individual meta-features and the level of feature interdependencies. The results underline the need to treat social bots as dynamic adversaries and have direct implications for the design, evaluation, and lifespan of bot-detection systems trained on historical behavioural features.

\end{abstract}

\noindent\textbf{Keywords:} social bots; Twitter; behavioural dynamics; time
series; stationarity; longitudinal analysis; bot detection.

\section{Introduction}

Automated accounts, or social bots, play an increasingly prominent role on
social media platforms. They are deployed at scale to amplify hashtags, push
links, promote commercial products, and shape political and health-related
narratives \citep{ferrara2016rise,cresci2020decade}. Concerns about their
impact have led to a large body of work on bot detection, much of which
constructs feature-based models from network, content, and account metadata
\citep{varol2017online,gilani2019largescale,wu2018twitter, loyola2019contrast,
heidari2021empirical}. 

Most of these systems treat behaviour as essentially static: features are
computed from a snapshot of activity and used to train classifiers that are
then assumed to remain valid over time. Yet both intuition and empirical work
suggest that bots respond to platform policies, detection tools, and evolving
user practices \citep{crescietal2017paradigm,pozzana2020measuring,lee2011sevenmonths}. If bot
operators adapt their strategies, then features that once reliably
distinguished bots from humans may drift, potentially degrading classifier
performance.

Understanding whether bots change their behaviour over time is therefore
crucial for both theory and practice. Prior work has examined bot and human
behavioural differences \citep{gilani2019largescale,varol2017online},
long-term spam campaigns and content polluters \citep{lee2011sevenmonths}, and
the dynamics of coordinated botnets \citep{pozzana2020measuring}. However, we
still know relatively little about how the behavioural signatures of labelled
promotional bots evolve over periods of many years.

Promotional bots---accounts that systematically advertise products, brands,
events, or services---are of particular interest. They rely heavily on URLs,
hashtags, and media to drive traffic, and often exploit sentiment, emojis, and
stylistic cues to appear more human-like and engaging
\citep{alrawi2020bots,puertas2019bots, loyola2019contrast,phan2020improving}. If such accounts
change how they post and what they post over time, then bot-detection models
that rely on static assumptions may quickly become outdated. In this paper we focus on a simple but fundamental question:

\medskip
\noindent\textbf{Research question.} \emph{Is there systematic evidence that
promotional social bots change their behaviour over time, and if so, along
which behavioural dimensions do these changes occur?}
\medskip

We operationalise this question in three parts:

\begin{itemize}
  \item RQ1: Are core behavioural meta-features of promotional bots
  (e.g.\ tweeting, URLs, hashtags, sentiment, emojis) stationary over time, or
  do they exhibit non-stationary trends?
  \item RQ2: Do behavioural meta-features differ systematically across
  \emph{generations} of bots introduced to the platform at different periods?
  \item RQ3: Do behavioural meta-features differ systematically across bots
  with different \emph{lifespans} (short-, mid-, and long-lived accounts)?
\end{itemize}

To answer these questions we build on a longitudinal dataset of promotional
Twitter bots collected between 2006 and 2021, focusing on a 12-year window
(2009--2020) where full-year coverage is available. Using ten
content-oriented meta-features extracted from 2.8 million tweets produced by
2{,}615 bot accounts, we construct yearly time series, perform stationarity
tests, and analyse trends across generations and age classes.

\subsection*{Contributions}

The work makes three contributions:

\begin{enumerate}
  \item We assemble and analyse a longitudinal dataset of promotional Twitter
  bots spanning 12 years, providing an empirical basis for studying the
  evolution of bot behaviour at scale.
  \item We demonstrate that common behavioural meta-features used in
  bot-detection research are non-stationary for promotional bots, and we
  quantify their deterministic and stochastic trends.
  \item We introduce generational and age-based stratifications of bot
  accounts, showing systematic differences between older and newer bots, and
  between short-lived and long-lived bots, across ten key behavioural
  dimensions.
  \item We explore the nature of associations between behavioural feature pairs, distinguishing
dependency from independence.
\item We analyse the dynamic changes in these relationships over time, including variations
in strength (weak, moderate, strong) and direction (increasing, decreasing,
or stable)

\end{enumerate}

Together, these findings establish a novel framework for understanding bot adaptation
strategies, providing a systematic approach to analysing how bots modify their
behaviours in response to changing environments over time. The results underline the need to treat social bots as dynamic adversaries whose behaviour changes over time, and they highlight concrete
dimensions along which this evolution can be observed.

\section{Background and Related Work}

Online social platforms have become key infrastructures for news consumption, political debate and everyday interaction. Alongside human users, they now host large numbers of automated and semi-automated accounts---commonly referred to as social bots---that generate, amplify, or curate content at scale \citep{ferrara2016rise}. Understanding how such bots behave, and whether their behaviour changes over time, is essential both for designing robust detection methods and for interpreting longitudinal studies based on historical platform data.

\subsection{Social bots and the detection arms race}

Social bots have been characterised as algorithmically controlled accounts that act within social media platforms while appearing, to varying degrees, human-like in their behaviour \citep{ferrara2016rise}. They range from benign automated news feeds and customer-service agents to coordinated disinformation systems and commercial spam networks. Taxonomies distinguish between fully automated bots, human-operated sockpuppets, and hybrid ``cyborg'' accounts that mix automation and human intervention \citep{pozzana2020measuring}. 

A large body of work has focused on detecting bots by exploiting features of user profiles, content, and network structure. Early approaches combined hand-crafted features (e.g., follower–followee ratios, posting frequency, URL usage) with classical machine-learning classifiers \citep{varol2017online}. Subsequent work has introduced more sophisticated feature sets, including content-based signals, temporal posting patterns, interaction networks, and graph embeddings \citep{gilani2019largescale}. Several surveys and systematic reviews provide comprehensive overviews of this rapidly evolving field, covering supervised and unsupervised techniques, traditional models and deep learning, and graph-based approaches \citep{orabi_detection_2020,cresci2020decade,aljabri_machine_2023,ellaky_systematic_2023}. Together, these studies depict bot detection as an ongoing arms race: as classifiers improve, bot designers adapt their tactics, which in turn motivates new detection strategies \citep{crescietal2017paradigm,cresci_coming_2021}.

\subsection{Behavioural characterisation of bots}

Beyond binary detection, a parallel strand of work seeks to understand how bots behave and how they differ from humans in aggregate. \citet{varol2017online} analysed millions of Twitter accounts and reported systematic differences in activity volume, content diversity, and connectivity between likely bots and humans. \citet{gilani2019largescale} compared large samples of bots and humans, showing that bots tend to post more frequently, share more URLs, and exhibit distinctive interaction patterns (e.g., retweeting and mentioning strategies) even when profile-level signals appear human-like.

Other studies have examined the structure and behaviour of specific botnets in detail. \citet{abokhodair_dissecting_2015} qualitatively analysed a long-lived Syrian political botnet, documenting its growth over time and its use of coordinated messaging to drown out genuine discussion. \citep{lee2011sevenmonths} followed ``content polluters'' on Twitter over seven months, describing how such accounts continuously polluted hashtags with promotional or irrelevant content. These studies provide rich descriptions of bot strategies and lifecycles, but they typically focus on particular events or campaigns.

A complementary line of work characterises bot behaviour using sequence- or pattern-based representations. \citep{crescietal2017paradigm} argued that modern ``social spambots'' emulate human profiles while exhibiting highly regular behavioural patterns over time, and proposed ``social fingerprinting'' techniques that encode action sequences (e.g., tweet/retweet/reply patterns) for group-level detection. \citet{pozzana2020measuring} modelled temporal dynamics of bots and humans using behavioural Markov chains, showing that bots and humans occupy different regions of a low-dimensional dynamical space and that short-term behavioural trajectories can be used to discriminate between them. Collectively, these works highlight that temporal regularities and meta-level action patterns are informative signals for both detection and analysis.

\subsection{Temporal dynamics, evolution, and concept drift}

The focus of most bot studies has been cross-sectional: they describe how bots differ from humans at a given time, or they evaluate classifiers on static datasets. However, there is growing evidence that bot behaviour and the broader bot ecosystem are not static. Lee et al.\ \citep{lee2011sevenmonths} documented long-term persistence of a population of content polluters, but did not systematically test whether their behavioural patterns changed over time. \citet{abokhodair_dissecting_2015} followed a political botnet for 35 weeks and showed how its size and topical focus evolved, but again the emphasis was on qualitative description rather than formal analysis of behavioural shifts.

From a detection perspective, concept drift - changes in the joint distribution of features and labels over time - poses a well-known challenge for machine learning \citep{gama_survey_2014}. In the context of social bots, \citet{echeverria_lobo_2018} introduced LOBO, a methodology that trains classifiers on one class of bots and evaluates them on previously unseen bot classes, showing that models that perform well on one dataset can fail dramatically on others. This suggests that bot populations differ across campaigns and time periods, and that fixed feature sets may not generalise robustly.

\citep{cresci2020decade} and \citet{cresci_coming_2021} argue that the bot–detector interaction should be understood as an adversarial process: as detectors heavily rely on specific behavioural or network features, bot designers have strong incentives to adapt and obfuscate those signals. Recent work on adversarial and deep learning-based bot detection further emphasises the risk that detectors overfit to particular datasets or behaviours \citep{aljabri_machine_2023,ellaky_systematic_2023}. These concerns are reinforced by temporal and synchrony-based detection systems such as DeBot \citep{chavoshi_debot_2016} and RTbust \citep{mazza_rtbust_2019}, which rely explicitly on fine-grained timing correlations and temporal patterns; if bots change how, when, and with whom they interact, the effectiveness of such methods may degrade.

Despite these insights, relatively few studies directly measure behavioural change in the same population of bots over extended periods. \citet{pozzana2020measuring} analyse short- to medium-term dynamics of behavioural trajectories for bots and humans, but they do not address whether the underlying feature distributions remain stable over years. Similarly, work on ``evolved spambots'' \citep{crescietal2017paradigm} demonstrates that newer bots can be engineered to evade detectors trained on older ones, without fully quantifying how their observable behavioural features have changed over time. This leaves open the question of whether, and to what extent, individual bots or bot populations systematically alter their observable behaviour as platforms, detection methods and campaign goals evolve.

\subsection{Data persistence, deletions, and longitudinal bias}

Longitudinal analyses of social media behaviour typically rely on historical datasets that are recollected or ``rehydrated'' through platform APIs. However, such datasets are affected by tweet and account deletions, which can introduce systematic biases. Almuhimedi et al.\ \citep{almuhimedi_tweets_2013} showed that a non-trivial fraction of tweets is deleted and that deletions are associated with specific kinds of content and user behaviour. Zubiaga and colleagues \citep{zubiaga_longitudinal_2018} conducted a longitudinal assessment of the persistence of multiple Twitter datasets, demonstrating that recollected corpora can lose a significant portion of their original content over time and that the surviving subset is not necessarily representative of the original population.

More recent work has examined deletions as a strategic tool. Torres-Lugo et al.\ \citep{torreslugo2022manipulating} analysed large-scale deletion patterns and showed that some accounts systematically delete tweets in ways that can game trending algorithms and evade retrospective scrutiny. Mubarak et al.\ \citep{mubarak_detecting_2023} proposed methods to predict tweet deletions and to identify likely reasons for deletion (e.g., abusive content, regret, or strategic removal), showing that deletion tendencies are themselves structured and predictable. These studies imply that longitudinal analyses based on recollected data may disproportionately capture particular behaviours and actors, while missing others that are more likely to delete content or be suspended.

For research on bots, data persistence is especially critical. Suspended or self-deleting bot accounts may vanish entirely from the platform, while surviving bots may selectively prune their timelines. If deletions are correlated with specific behaviours (e.g., high-volume spam, certain topics, or aggressive interaction patterns), then longitudinal samples constructed at later dates may over-represent bots with more ``cautious'' or human-like behaviours and under-represent those whose content has been removed. This has direct implications for studies that attempt to assess how bots behave over time using rehydrated data.

\subsection{Summary and research gap}

Existing work thus provides a rich picture of what social bots are, how they can be detected, and how they differ from humans in terms of activity, content, and network position \citep{ferrara2016rise,varol2017online,gilani2019largescale,cresci2020decade,aljabri_machine_2023}. There is also clear evidence that bot ecosystems evolve, that detection methods may fail to generalise to new bot classes, and that historical datasets are affected by deletions and suspensions \citep{echeverria_lobo_2018,crescietal2017paradigm,almuhimedi_tweets_2013,zubiaga_longitudinal_2018,torreslugo2022manipulating}. However, there remains a specific gap around the question addressed in this study:

\begin{quote}
\emph{Is there empirical evidence that social bots systematically change their observable behaviour over time, when measured using widely employed behavioural and meta-data features, in longitudinal samples of the same bot accounts?}
\end{quote}

Most prior work touches on temporal aspects either indirectly (through cross-dataset generalisation tests or short-term dynamics) or within specific case studies without formal tests of behavioural change. Few studies repeatedly sample the same population of known bots over extended periods, quantify changes in the distribution of common behavioural features, and explicitly test whether these features can be treated as stationary for the purposes of modelling and detection. 

The remainder of this paper addresses this gap by examining labelled bot accounts across multiple time slices and analysing whether key behavioural indicators---such as posting volume, interaction patterns, and content usage---exhibit significant shifts over time or remain broadly stable within the limits imposed by data persistence and deletion.

\section{Data and Methods}\label{sec:methods}

\subsection{Research design}

We adopt a two-stage research design. First, we treat the yearly counts of
specific behavioural meta-features as time series and apply standard
stationarity tests to assess whether these features can be modelled as
stationary processes or whether they exhibit systematic trends. Second, we
stratify bot accounts by generation and age class and examine how the
distribution of feature values varies across these strata.

This design directly addresses RQ1 (stationarity), while the stratified
analyses address RQ2 (generational differences) and RQ3 (age differences).

\subsection{Data collection}

We focus on promotional spambots active on Twitter between 2006 and 2021.
Following prior work, we start from ground-truth bot IDs in publicly available
research corpora, including multiple datasets from the MIB project and
Botometer's Bot Repository. These datasets primarily comprise promotional and
spammer accounts and have been widely used in bot-detection research.

Using the Twint Python library, we scraped tweets from these bot accounts.
Twint collects historical tweets without relying on the official Twitter API,
which simplifies large-scale retrieval and avoids rate limits. For each
account, we collected up to 3{,}200 of the most recent tweets, along with
associated metadata.

In total, we obtained 2{,}798{,}672 tweets from 2{,}615 unique bot accounts.
Table~\ref{tab:dataset} summarises the constituent corpora and their coverage.
Although raw data extend from 2006 to early 2021, we restrict our analysis to
the period 2009--2020 in order to work with complete calendar years and to
avoid gaps caused by collection interruptions.

\begin{table}[t]
\centering
\caption{Summary of collected dataset of promotional bots.}
\label{tab:dataset}
\begin{tabular}{lrrr}
\toprule
Dataset & \# Accounts & \# Tweets & Covered years \\
\midrule
traditional spambots \#1 & 565 & 350{,}512 & 2007--2021 \\
traditional spambots \#3 & 104 & 220{,}582 & 2009--2021 \\
traditional spambots \#4 & 227 & 96{,}109 & 2013--2021 \\
social spambots \#2     & 694 & 104{,}813 & 2014--2020 \\
social spambots \#3     & 151 & 1{,}084{,}711 & 2012--2021 \\
caverlee\_2011          & 205 & 535{,}155 & 2006--2021 \\
gilani\_2017            & 33  & 90{,}138  & 2016--2021 \\
varol\_2017             & 37  & 87{,}874  & 2008--2021 \\
cresci\_stock\_2018     & 39  & 72{,}804  & 2016--2021 \\
fake\_followers         & 560 & 155{,}974 & 2012--2019 \\
\midrule
\textbf{Total}          & \textbf{2{,}615} & \textbf{2{,}798{,}672} & 2007--2021 \\
\bottomrule
\end{tabular}
\end{table}

\subsection{Behavioural meta-features}

We focus on ten content-based behavioural meta-features that have been widely
used in bot-detection research and that are particularly relevant to
promotional activity. To minimise bias, we selected features that have been
validated in prior studies as useful for distinguishing bots from humans or
for characterising automated campaigns
\citep{wu2018twitter,sedhai2017analysis,loyola2019contrast,phan2020improving,puertas2019bots,alrawi2020bots}.

For each bot account and year, we compute the following meta-features:

\begin{enumerate}
  \item \textbf{Tweeting.} Count of original tweets posted by the account.
  \item \textbf{Retweeting.} Count of retweets.
  \item \textbf{Replying.} Count of reply tweets.
  \item \textbf{URLs.} Count of tweets that contain at least one URL.
  \item \textbf{Duplicated text.} Count of tweets whose cleaned textual content
  is identical to another tweet from the same account (a proxy for template
  use and repetitive messaging).
  \item \textbf{Hashtags.} Count of tweets that contain at least one hashtag.
  \item \textbf{Sentiment.} Count of tweets expressing positive, negative, or
  neutral sentiment, based on a lexicon-based sentiment analysis. Following
  prior work, we treat scores below $-0.2$ as negative, above $0.2$ as
  positive, and between $-0.2$ and $0.2$ as neutral, using thresholds that
  have been empirically validated for social media text
  \citep{phan2020improving,monica2020fake}.
  \item \textbf{Languages.} Number of distinct ISO~639 language codes observed
  in an account's tweets in a given year.
  \item \textbf{Emojis.} Count of tweets containing at least one emoji.
  \item \textbf{Media.} Count of tweets including at least one photo or video.
\end{enumerate}

These meta-features aggregate over the raw tweet-level features listed in
Table~\ref{tab:features-raw}. For example, the tweeting, replying, and
retweeting meta-features are derived from action types at the tweet level,
while duplicated text is derived from pairwise comparisons of cleaned text
strings across an account's tweets.

\begin{table}[t]
\centering
\caption{Tweet-level behavioural features used to construct meta-features.}
\label{tab:features-raw}
\begin{tabular}{ll}
\toprule
Category & Values \\
\midrule
Posting action & tweet, retweet, reply \\
URL           & contains URL / no URL \\
Sentiment     & positive / negative / neutral \\
Media         & at least one image / at least one video / none \\
Timestamp     & year, month, day of tweet \\
Text          & duplicated / non-duplicated / empty \\
Hashtag       & contains hashtag / no hashtag \\
Emoji         & contains emoji / no emoji \\
Language      & ISO~639 language codes (e.g.\ en, es, zh, \dots) \\
\bottomrule
\end{tabular}
\end{table}

\subsection{Temporal aggregation}

For each meta-feature, we aggregate counts at yearly resolution across all bots
in the dataset for the period 2009--2020. This yields one time series per
meta-feature, with 12 annual observations. The year 2009 corresponds to the
earliest year for which we have sufficient complete data across the constituent
datasets; 2020 is the latest complete year before collection interruptions in
2021.

We also construct per-feature distributions for two stratifications that we
use in later analyses:

\begin{itemize}
  \item \textbf{Generations.} Bots are grouped into three equally sized
  \emph{generations} based on the year in which they first appear in our
  data: first generation (2009--2012), second generation (2013--2016), and
  third generation (2017--2020).
  \item \textbf{Age classes.} Bots are grouped into three \emph{age classes}
  based on the length of their observed lifespan in the dataset: short-lived
  (1--4 years), mid-lived (5--8 years), and long-lived (9--12 years).
\end{itemize}

These groupings allow us to compare behavioural patterns across cohorts of bots
with different activation periods and lifespans.

\subsection{Stationarity analysis}

To address RQ1, we test whether each meta-feature time series is stationary.
We use two complementary tests:

\begin{itemize}
  \item The \textbf{Augmented Dickey--Fuller (ADF)} test
  \citep{mushtaq2011adf} evaluates the null hypothesis that a time series has
  a unit root (i.e.\ is non-stationary) against the alternative of
  stationarity. Failure to reject the null suggests non-stationarity.
  \item The \textbf{Kwiatkowski--Phillips--Schmidt--Shin (KPSS)} test
  \citep{syczewska2010kpss} evaluates the null hypothesis of stationarity
  against the alternative of a unit root. We use both the level and trend
  variants of the test to distinguish different forms of non-stationarity.
\end{itemize}

For each meta-feature we estimate the ADF statistic with an appropriate lag
order (selected via standard information criteria) and report the associated
$p$-value. We use 5\% as the primary significance level. Similarly, for KPSS we
compute test statistics for level- and trend-stationarity and record the
corresponding $p$-values.

Combining these tests allows us to characterise each series qualitatively as:

\begin{itemize}
  \item \emph{Stationary} if we reject the ADF null (no unit root) and fail to
  reject the KPSS null (stationary).
  \item \emph{Trend-stationary (deterministic trend)} if the series becomes
  stationary after detrending and KPSS indicates trend-stationarity.
  \item \emph{Stochastic trend (unit root)} if both tests suggest the presence
  of a unit root, indicating a random-walk-like process.
\end{itemize}

\subsection{Trend estimation and predictability}

For each non-stationary meta-feature, we estimate a linear trend over the
12-year period using ordinary least squares regression:

$
  \text{Count}_{t} = m \cdot t + b,
$

where $t$ indexes years (2009, \dots, 2020), $m$ is the slope, and $b$ is the
intercept. A positive slope indicates an upward trend (increasing counts over
time), while a negative slope indicates a downward trend. We use the magnitude
of the slope to compare relative rates of change across features.

Combining the KPSS classification with the slope provides an interpretable
characterisation of each feature's dynamics:

\begin{itemize}
  \item Features with a deterministic trend and large slope are considered
  highly predictable, with consistent directional change over time.
  \item Features with a stochastic trend are treated as less predictable and
  more sensitive to external shocks and idiosyncratic fluctuations.
\end{itemize}

\subsection{Generational and age-based analysis}

To address RQ2 and RQ3, we conduct descriptive comparisons of meta-feature
distributions across generations and age classes. For each group we compute
aggregate statistics such as mean and median counts per account, proportions
(e.g.\ percentage of tweets containing URLs or hashtags), and distributional
summaries for textual and expressive features (duplicated text, sentiment,
emojis, languages).

Visualisations (time-series plots and small multiples of distribution plots)
are used to explore patterns, but in this paper we report the qualitative
patterns and quantitative summaries derived from these underlying analyses.

\section{Results}

\subsection{Non-stationarity of behavioural meta-features (RQ1)}

\begin{figure}[t]
  \centering
  \includegraphics[width=\textwidth]{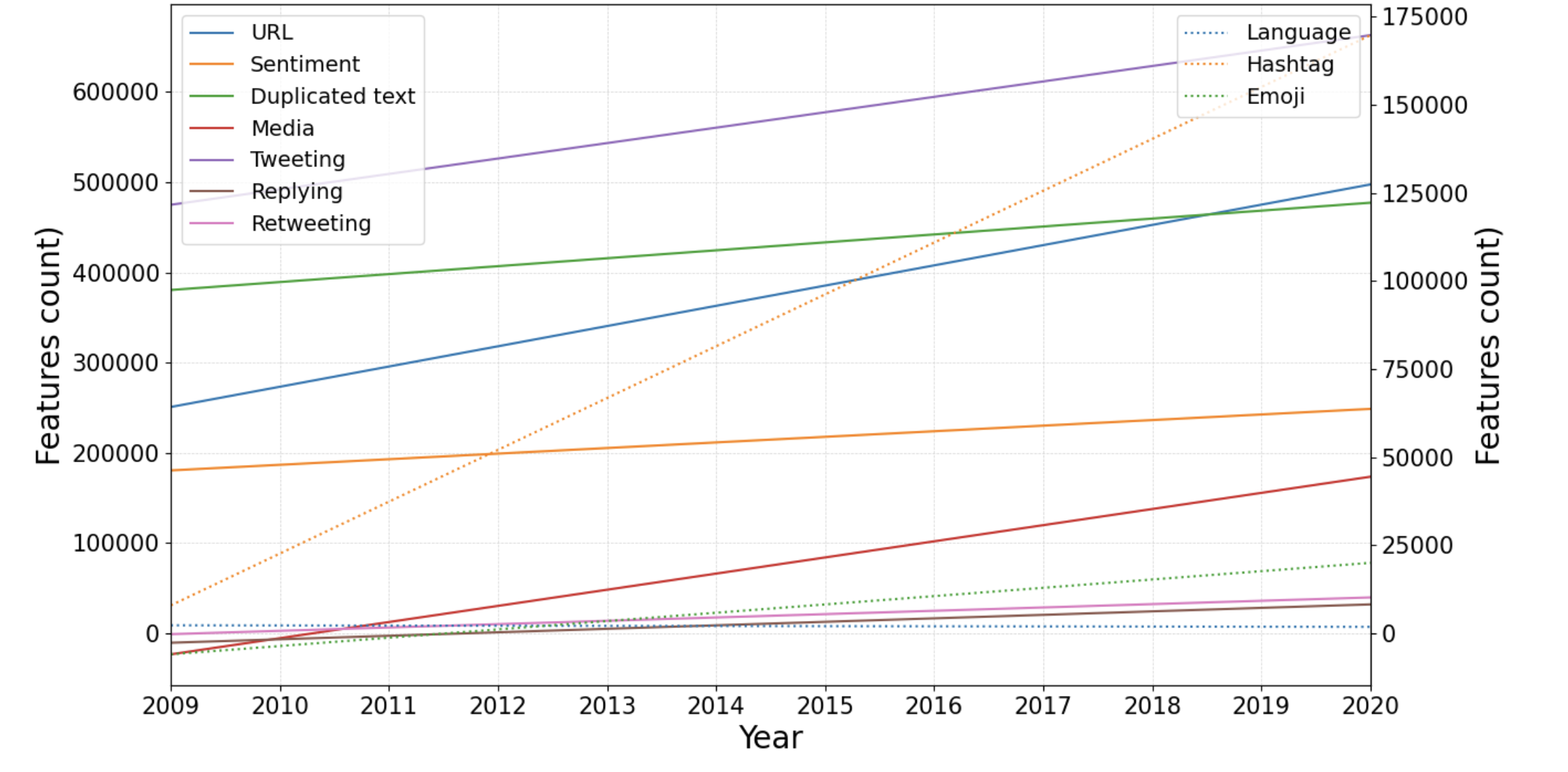}
  \caption{Yearly time series (2009--2020) of the ten behavioural
  meta-features for promotional bots: tweeting, retweeting, replying, URLs,
  hashtags, duplicated text, sentiment, languages, emojis, and media. Each
  panel shows the total yearly count across all bots for the corresponding
  meta-feature.}
  \label{fig:time-series}
\end{figure}

Figure~\ref{fig:time-series} shows the yearly time series (2009--2020) for the
ten behavioural meta-features considered in this study. Visual inspection
suggests that most features exhibit clear upward trends, with the exception of
language diversity, which appears to decline slightly. In particular, tweeting
activity, retweeting, replying, URLs, hashtags, duplicated text, sentiment,
emojis, and media all increase over the 12-year period.

To formalise these observations, we apply the Augmented Dickey--Fuller (ADF)
test to each meta-feature time series. Table~\ref{tab:adf} summarises the ADF
statistics and $p$-values. For all ten meta-features the ADF test fails to
reject the null hypothesis of a unit root at the 5\% significance level
($p>0.05$ in all cases), indicating that none of the series can be treated as
stationary over the period 2009--2020.

\begin{table}[t]
\centering
\caption{Augmented Dickey--Fuller test results for bot behavioural
meta-features (2009--2020). Higher statistics and larger $p$-values indicate
non-stationarity.}
\label{tab:adf}
\begin{tabular}{lrrr}
\toprule
Meta-feature & ADF statistic & Lag & $p$-value \\
\midrule
Tweeting        &  1.63 & 2 & 0.714 \\
Retweeting      & -2.54 & 2 & 0.368 \\
Replying        & -0.83 & 3 & 0.811 \\
URLs            & -1.42 & 2 & 0.794 \\
Hashtags        & -0.68 & 2 & 0.960 \\
Emojis          & -1.09 & 2 & 0.907 \\
Media           & -2.43 & 2 & 0.410 \\
Sentiment       & -1.59 & 2 & 0.730 \\
Duplicated text & -1.65 & 2 & 0.707 \\
Languages       & -1.83 & 2 & 0.636 \\
\bottomrule
\end{tabular}
\end{table}

Complementary Kwiatkowski--Phillips--Schmidt--Shin (KPSS) tests clarify the
type of non-stationarity. For each meta-feature we compute the KPSS statistics
for level- and trend-stationarity. Table~\ref{tab:kpss} summarises the
resulting $p$-values and classifies each series as either following a
deterministic trend or a stochastic trend.

\begin{table}[t]
\centering
\caption{KPSS results and linear trend slopes for behavioural meta-features.
The slope is the coefficient of a linear regression of yearly counts on time
(2009--2020); positive slopes indicate upward trends, negative slopes downward
trends.}
\label{tab:kpss}
\begin{tabular}{lrrrrl}
\toprule
Meta-feature & Level $p$ & Trend $p$ & Trend type & Slope & Direction \\
\midrule
URLs            & 0.10   & 0.090 & deterministic & 22{,}413 & upward \\
Media           & 0.094  & 0.086 & deterministic & 17{,}889 & upward \\
Tweeting        & 0.10   & 0.10  & deterministic & 17{,}084 & upward \\
Hashtags        & 0.069  & 0.10  & deterministic & 14{,}712 & upward \\
Duplicated text & 0.10   & 0.10  & deterministic &  8{,}790 & upward \\
Sentiment       & 0.10   & 0.10  & deterministic &  6{,}196 & upward \\
Replying        & 0.085  & 0.064 & deterministic &  3{,}866 & upward \\
Retweeting      & 0.036  & 0.010 & stochastic    &  3{,}698 & upward \\
Emojis          & 0.049  & 0.063 & deterministic &  2{,}361 & upward \\
Languages       & 0.10   & 0.10  & deterministic &    -41   & downward \\
\bottomrule
\end{tabular}
\end{table}

For all features except retweeting, the KPSS results are consistent with a
deterministic trend: after detrending, the residuals are stationary. In
contrast, retweeting shows evidence of a stochastic trend, suggesting a
random-walk-like process with greater sensitivity to exogenous shocks (such as
events or platform-level changes).

F
Overall, the combination of ADF and KPSS tests and the observed trends provides
strong evidence that promotional bots do \emph{not} behave in a stationary way
over time. Instead, their activity levels and content characteristics change
systematically, with most meta-features exhibiting monotonic trends.

\subsection{Trends in behavioural features}

The slopes reported in Table~\ref{tab:kpss} and illustrated in
Figure~\ref{fig:time-series} allow a more detailed comparison of how quickly
different aspects of bot behaviour are changing.

The fastest-growing features are URLs and media. The yearly count of tweets
containing URLs increases with a slope of approximately 22{,}400, while the
count of tweets containing media (photos or videos) grows at around 17{,}900
tweets per year. Tweeting and hashtag usage also display substantial positive
slopes (17{,}084 and 14{,}712 respectively), indicating that bots are both more
active and more aggressive in using visibility-enhancing mechanisms.

Duplicated text, sentiment, replying, and emojis grow at more moderate rates.
The increase in duplicated text counts suggests that template-based messaging
remains a core strategy, even as bots diversify their content. The upward trend
in sentiment and emojis indicates that bots increasingly use affective cues and
expressive markers, moving away from purely neutral or functional messages.

In contrast, language diversity exhibits a small but consistent downward trend:
bots gradually concentrate their activity into a smaller set of languages,
often dominated by English. This may reflect a strategic focus on specific
markets or an attempt to better blend into the dominant linguistic environment
on the platform.

Taken together, these trends support the view that promotional bots have become
more active, more content-rich, and more expressive over time, while narrowing
their linguistic focus.

\subsection{Behaviour across generations (RQ2)}

To examine generational differences, we group bots into three activation cohorts
based on the year in which they first appear in our dataset (i.e.\ the first
year for which we observe tweets from the account):

\begin{itemize}
  \item \textbf{First generation} (2009--2012)
  \item \textbf{Second generation} (2013--2016)
  \item \textbf{Third generation} (2017--2020)
\end{itemize}

\begin{figure}[htb!]
  \centering
  \includegraphics[width=0.9\textwidth]{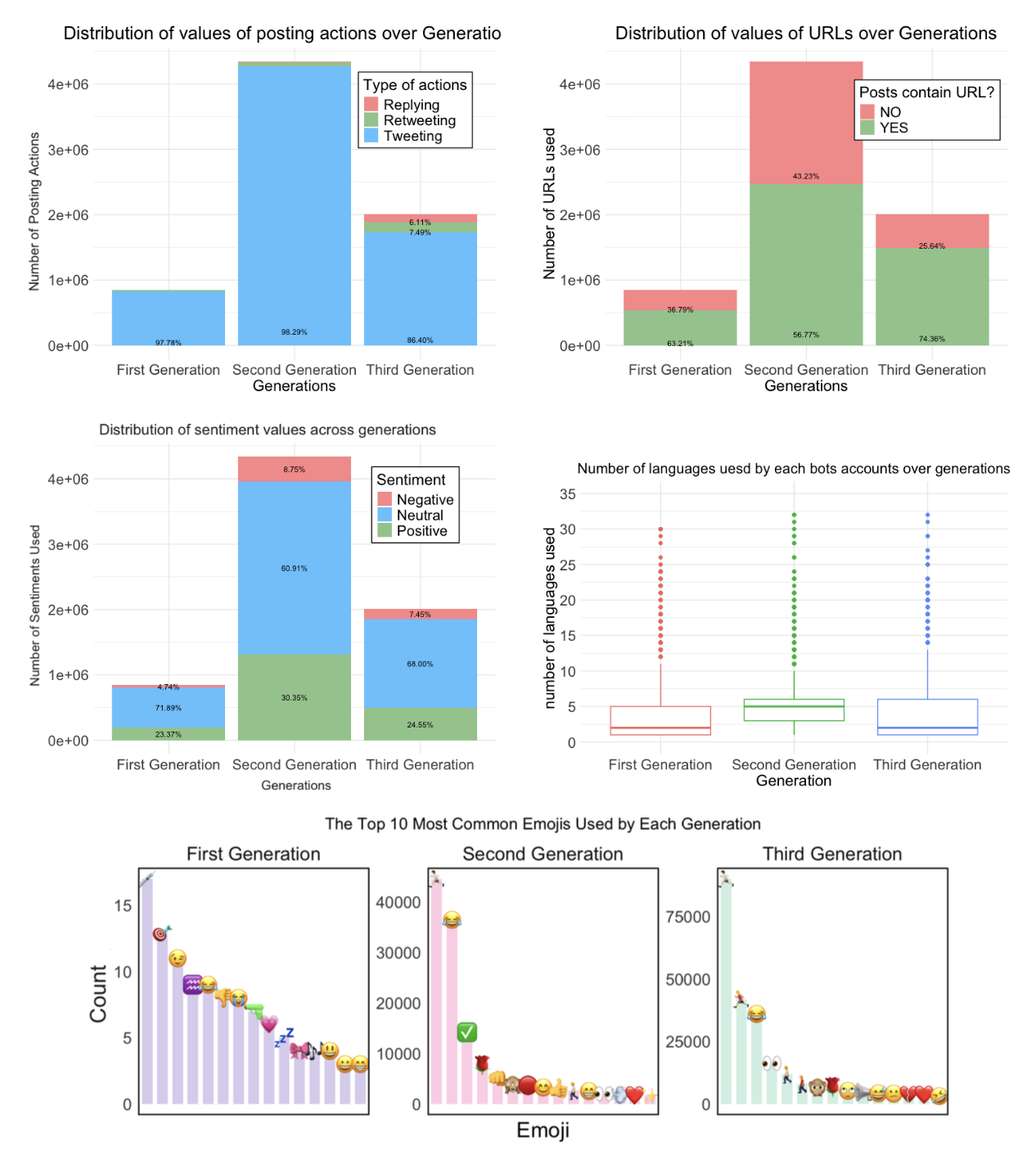}
  \caption{Behavioural meta-feature distributions for three generations of
  promotional bots. Each panel shows the distribution of a meta-feature
  (posting actions, URLs, sentiment, languages, emojis) for first, second, and third-generation bots.}
  \label{fig:generations}
\end{figure}

\begin{figure}[htb!]
  \centering
  \includegraphics[width=0.9\textwidth]{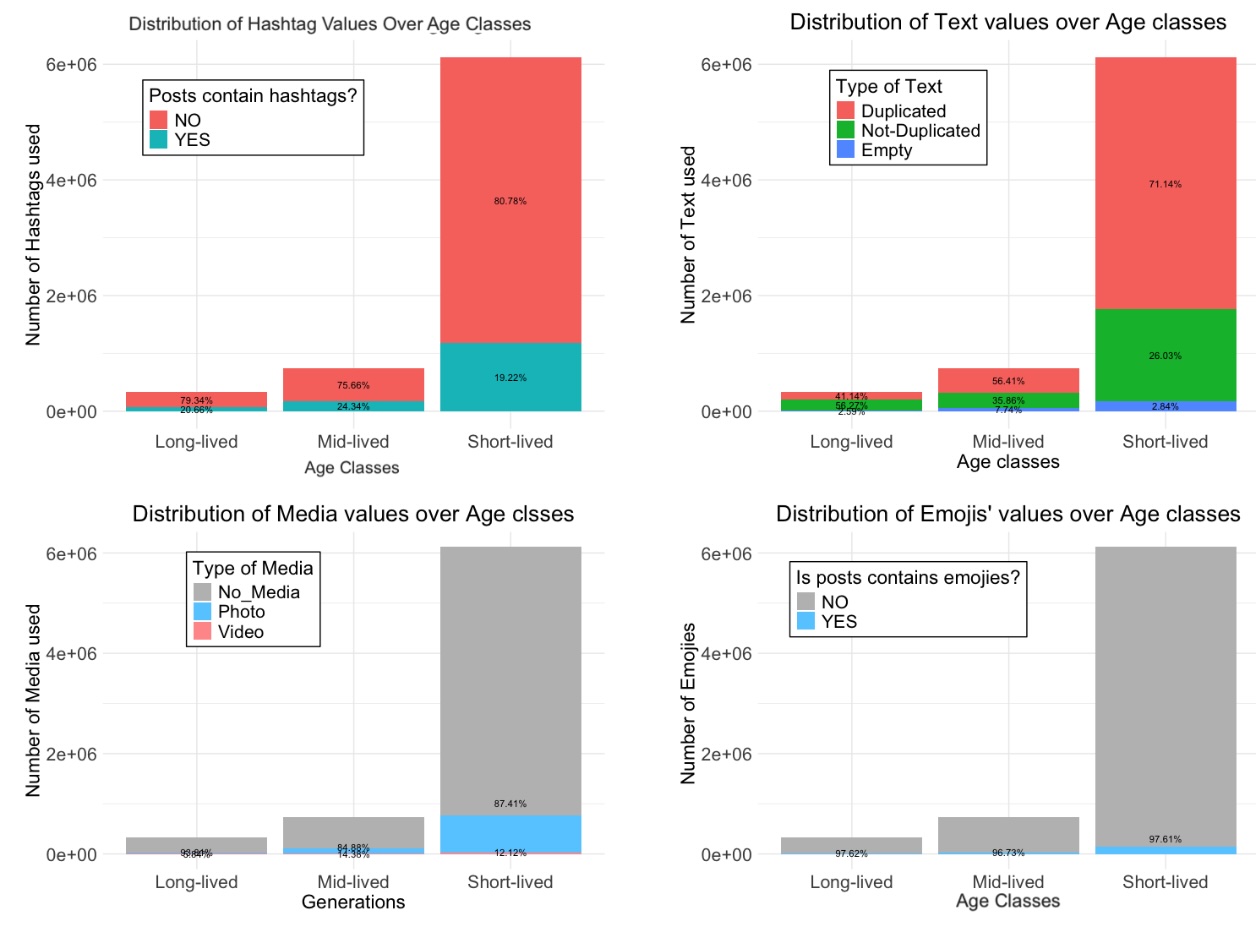}
  \caption{Further behavioural meta-feature distributions for three generations, showing 
  hashtags, text, media use, and emoji use, for first, second, and third-generation bots.}
  \label{fig:further-generations}
\end{figure}

For each generation we compute the distribution of all behavioural meta-features
defined in Section~\ref{sec:methods}, and compare how posting actions, URLs,
sentiment, text duplication, emojis, languages, media, and hashtags differ across
cohorts. Figures~\ref{fig:generations} and \ref{fig:further-generations} summarise these distributions; the
paragraphs below describe the main patterns.

\paragraph{Posting actions.}

Across all generations, tweeting is the dominant posting action, followed by
retweeting and then replying. However, there are clear generational shifts in
the overall intensity and composition of these actions.

The first generation has the lowest average number of posting actions (tweets,
retweets, replies) per account. Activity is present but relatively modest, and
replies are rare, indicating predominantly broadcast-oriented behaviour. The
second generation has the highest average number of posting actions, reflecting
a period of very intensive, campaign-like activity. The third generation sits
between the first and second generations in terms of overall volume, but with
richer and more varied behaviour.

Replying only begins to appear as a substantial action pattern in the second
generation. In the first generation, bots are overwhelmingly broadcast-oriented:
they tweet and retweet but seldom reply. In the second and third generations,
replying becomes more common, indicating that newer bots are more likely to
mimic conversational behaviour (e.g.\ thanking users, responding to mentions)
rather than simply pushing content.

\paragraph{URLs.}

URLs are central to promotional bot behaviour in all three generations, but
their prevalence and intensity change over time. As overall posting volume rises
from the first to the second generation, the average number of posts containing
URLs also increases.

The second generation exhibits the highest average number of URLs per account.
In this cohort, posts containing URLs form the majority of all posts, with bot
posts containing URLs constituting approximately 56.8\% of all actions. Across
all generations, the ratio of posts containing URLs is higher than the ratio of
posts without URLs, underscoring the central role of link sharing in
promotional campaigns.

These results indicate that each successive generation of bots relies heavily on
URLs to drive traffic to external sites (e.g.\ job pages, product pages,
landing pages), with the second generation in particular exploiting URLs at
very high rates.

\paragraph{Sentiment.}

The distribution of sentiment is broadly stable across generations. In all three
cohorts, neutral sentiment dominates, followed by positive and then negative
sentiment. This basic ranking is preserved, suggesting that baseline emotional
tone is similar across generations despite other behavioural changes.

However, the relative proportions of positive and negative tweets show a slight
increase over time. In later generations, bots contribute a higher share of
emotionally charged content (both positive and negative) and a slightly lower
share of strictly neutral content. This supports the idea that newer
promotional bots increasingly exploit emotional language---for example, using
enthusiastic or urgent phrasing in product promotions---to engage audiences and
appear more human-like.

\paragraph{Text and duplication.}

Text duplication is a prominent characteristic of promotional bots in every
generation. Across all three cohorts, the majority of posts contain duplicated
text: there is a repeated, cleaned text segment that appears in multiple posts
(often across many accounts). This pattern is consistent with template-based
messaging and bulk posting.

The absolute amount of duplicated text increases with the total number of
postings, so the second generation---which has the highest posting volume---also
exhibits the largest volume of duplicated text. Unique text (non-duplicated
content) forms the second-largest category, while empty text (e.g.\ media-only
posts or posts with no significant textual content) is the smallest.

Over time there is a slight increase in both unique and empty text. This
suggests that, while templates remain central, later generations of bots mix
template-based posts with a somewhat greater amount of idiosyncratic or media-
heavy content, making them appear less mechanical and more varied.

\paragraph{Emojis.}

Emojis are relatively rare compared to plain text, but their usage increases and
becomes more diverse over generations.

Across all generations, the majority of posts do not contain emojis.
Nevertheless, the average number of emojis per post increases slightly from one
generation to the next, and the diversity of emojis used also grows.

\begin{itemize}
  \item \emph{First generation.} Emoji usage is low: where emojis appear, posts
  typically contain around one emoji. The most common emojis are symbols
  representing specific ideas, concepts, or cultural references rather than
  explicitly emotional expressions.
  \item \emph{Second generation.} Emoji usage becomes more expressive and
  interaction-focused. Posts commonly contain between one and five emojis, and
  there are notable outliers where individual bots use extremely high numbers
  of emojis. Second-generation bots use more faces, hearts, and
  interaction-related symbols, suggesting a deliberate attempt to make posts
  more engaging and human-like.
  \item \emph{Third generation.} Emoji usage remains frequent (typically one to
  four emojis per emoji-containing post), again with significant outliers. Newer
  facial expressions appear, including more nuanced emotional emojis, indicating
  a further evolution towards fine-grained emotional expression.
\end{itemize}

In addition to these generational patterns, there is a core set of emojis that
is used across all three generations, providing continuity in expressive style.
Overall, the trajectory from the first to the third generation is one of
increasing expressive richness and a shift from symbolic to more emotional and
interactional emoji use.

\paragraph{Languages.}

Language diversity also varies by generation. The second generation of bot
accounts has the highest average number of languages used per account, with a
mean of 4.41 distinct languages and a wide range from 1 to 31 languages per
account.

The first and third generations are somewhat more restrained and similar to
each other, with average language counts of 3.91 and 3.08 languages per account,
respectively. These cohorts still contain multilingual bots, but on average
their language repertoire is slightly narrower than that of the second
generation.

This pattern suggests that the second generation includes many bots that
broadcast the same promotional content in multiple languages, possibly to reach
different national markets or user communities. In contrast, first- and
third-generation bots tend to be more focused on a smaller number of languages,
often dominated by English.

\paragraph{Media (images and video).}

Most posts made by bots across all generations do not contain any media: plain
text (with or without URLs and hashtags) is still the dominant format. However,
there is a clear upward trend in media usage.

The average number of images per account increases substantially from one
generation to the next. Early bots use images relatively sparingly; by the
second and especially the third generation, images are used much more
frequently, typically to advertise products, illustrate job opportunities, or
make promotional content visually attractive.

Videos are effectively absent in the first generation. They appear for the
first time in the second generation, and their usage increases slightly again in
the third generation. The introduction and gradual growth of video content
align with broader platform affordances and norms, and it shows that newer bots
adopt richer media to increase engagement.

\paragraph{Hashtags.}

Hashtag usage displays both continuity and change. At a coarse level, the
majority of posts in all generations do not contain any hashtags. When hashtags
are present, however, their number and thematic content reveal interesting
generational shifts.

In terms of volume, the number of hashtags used in posts varies significantly
from one generation to the next:

\begin{itemize}
  \item First generation: between 1 and 57 hashtags per hashtagged post.
  \item Second generation: between 1 and 59 hashtags per hashtagged post.
  \item Third generation: between 1 and 206 hashtags per hashtagged post.
\end{itemize}

The third generation therefore exhibits the most extreme hashtagging behaviour,
with some posts containing very large numbers of hashtags, consistent with
aggressive visibility-seeking strategies.

An analysis of the top 30 most frequent hashtags in each generation (reported in
the Appendix) shows both shared themes and cohort-specific emphases. Across all
generations, there is a strong focus on employment and human-resources topics
(e.g.\ \emph{job}, \emph{hiring}, \emph{career}). Each generation also reflects
the dominant technologies and interests of its time, with first-generation bots
highlighting generic IT and tech terms, second-generation bots emphasizing
platforms such as \emph{Android} and \emph{SoundCloud}, and third-generation
bots promoting more recent technologies and devices.

\medskip

Taken together, these results show that bot behaviour differs systematically
across generations. Newer cohorts are more active, more media- and emoji-rich,
often more multilingual, and employ more aggressive and topical hashtagging
than earlier cohorts. This provides further evidence that social bots are not
static entities: their behavioural signatures evolve over time in ways that
reflect both changing platform affordances and adaptive strategies by bot
operators.

\subsection{Behaviour across age classes (RQ3)}

We now compare behavioural patterns across three account age classes, defined by
the length of the observed lifespan of each bot in the dataset:

\begin{itemize}
  \item \textbf{Short-lived} bots: active for 1--4 years.
  \item \textbf{Mid-lived} bots: active for 5--8 years.
  \item \textbf{Long-lived} bots: active for 9--12 years.
\end{itemize}

\begin{figure}[htb!]
  \centering
  \includegraphics[width=0.9\textwidth]{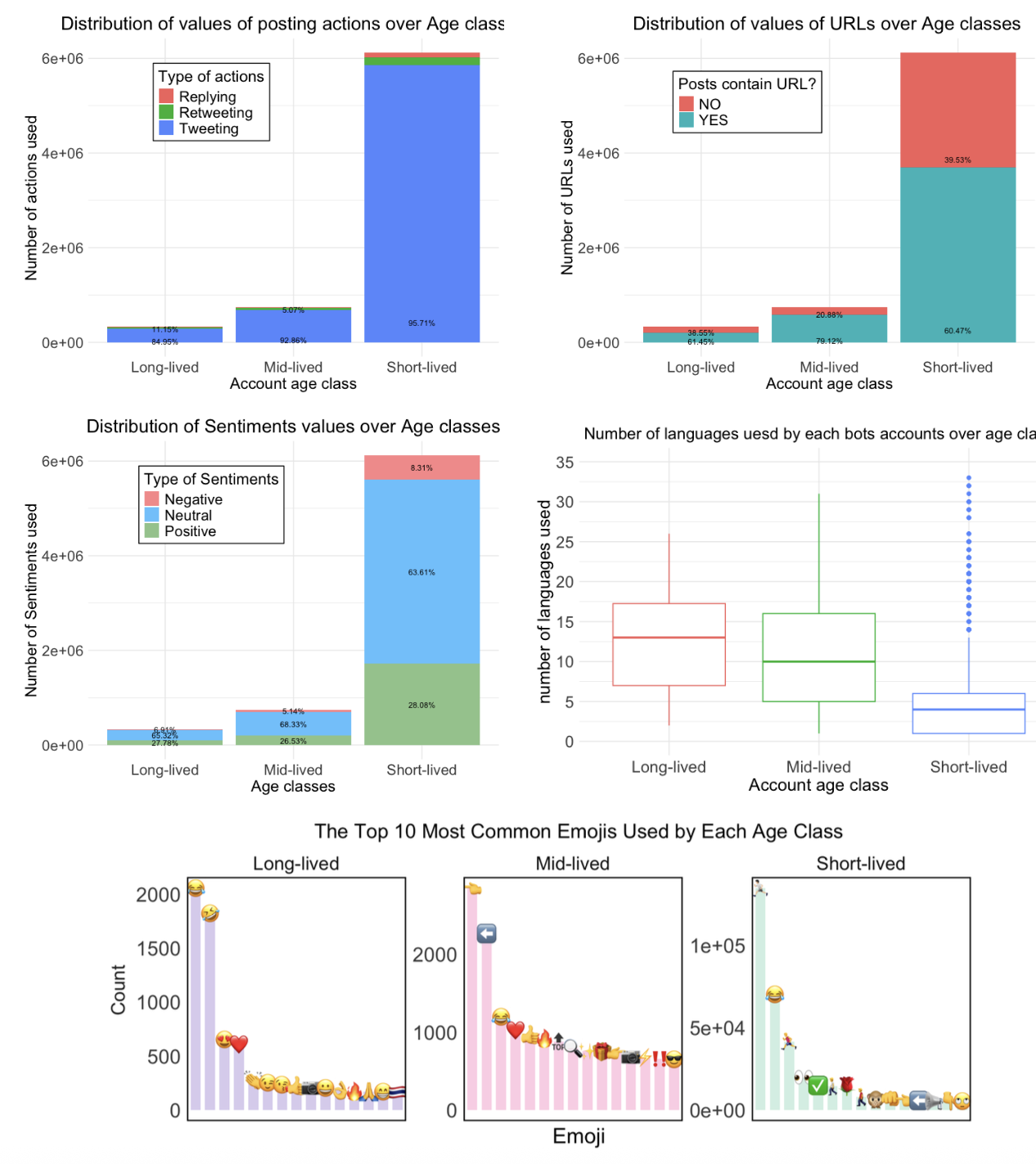}
  \caption{Behavioural meta-feature distributions for three age classes of
  promotional bots: short-lived (1--4 years), mid-lived (5--8 years), and
  long-lived (9--12 years). Each panel shows the distribution of a
  meta-feature for the corresponding age class.}
  \label{fig:age-classes}
\end{figure}

For each age class we compute the distribution of the same behavioural
meta-features as before. Figure~\ref{fig:age-classes} summarises these
distributions; the main differences are discussed below.

\paragraph{Posting actions and duplicated text.}

Short-lived bots exhibit the highest average number of posting actions in their
timelines. Their activity is intense but concentrated in a relatively short
period, consistent with bots deployed for specific campaigns (e.g.\ product
launches or political events). They also post the largest proportion of
duplicated text; in some cases, more than 70\% of their tweets are near
verbatim copies of earlier messages, indicating heavy use of templates.

Long-lived bots, in contrast, show lower posting rates but are more evenly
distributed over time. Their duplicated text proportion is still substantial,
but lower than that of short-lived bots. This suggests that longevity is
associated with less extreme and more varied posting behaviour, potentially
helping these bots evade detection.

\paragraph{URLs and hashtags.}

Both short-lived and long-lived bots have a high proportion of tweets containing
URLs (around 60\% of posts), but mid-lived bots have the highest proportion
(over 79\%). Hashtag use follows a similar pattern, with short-lived bots
combining intense posting with heavy hashtag usage, again consistent with
campaign-style behaviour.

\paragraph{Sentiment.}

All three age classes show similar sentiment distributions: approximately
two-thirds of tweets are neutral, around 28\% are positive, and around 7\% are
negative. This suggests that, while activity levels and strategies differ,
baseline sentiment distributions are relatively stable across lifespan.

\paragraph{Emojis, languages, and media.}

Short-lived bots use the most emojis on average and show a wide range of emoji
counts per tweet, including some extreme outliers with very large numbers of
emojis. They frequently employ celebratory and attention-grabbing emojis, as
well as approval markers such as check marks and eye symbols.

Long-lived bots, by contrast, use a moderate number of emojis but display a
wider variety of emotional states (e.g.\ emojis for heat, cold, appetite, or
pleading), suggesting more nuanced affective expression. Mid-lived bots fall
between these extremes but tend to employ a rich variety of positive-emotion
emojis.

Language diversity increases with account age: short-lived bots have relatively
few languages, mid-lived bots more, and long-lived bots the most. This suggests
that longer-lived bots engage with a broader set of linguistic communities,
perhaps reflecting strategic expansion over time.

Across all age classes, the majority of tweets contain no media (around 89\%),
but the remaining posts with photos and videos play an important role in making
promotional content engaging and credible.

\subsection{Predictability of features}

Combining the trend classifications and slopes, we can characterise the
predictability of bot behavioural features.

Highly predictable features include URLs, media, tweeting, hashtags, duplicated
text, sentiment, replying, emojis, and languages. These features follow
deterministic trends and change in a consistent, monotonic fashion over time.
Retweeting is less predictable: despite an upward slope, its stochastic trend
suggests greater sensitivity to exogenous events and campaign-specific
dynamics. Bots may retweet opportunistically in response to trending topics and
real-time events, leading to more irregular behaviour.

Overall, the results across RQ1--RQ3 provide strong evidence that promotional
bots change their behaviour over time and that these changes are systematic
rather than random. Bot behaviour evolves across generations and account
lifespans, and core meta-features commonly used in detection models exhibit
clear non-stationary patterns.

\section{Discussion}

\subsection{Evidence of evolving bot behaviour}

The results provide clear, quantitative evidence that promotional bots do not
behave in a stationary way over time. Core behavioural meta-features that are
commonly used in detection---tweeting frequency, link sharing, hashtag usage,
duplicated text, sentiment, emojis, and media---all display non-stationary,
mostly deterministic trends. Bots have become more active, more media-rich,
and more expressive, while gradually narrowing their language repertoire.

These findings align with earlier qualitative characterisations of an arms
race between bot developers and detection systems
\citep{ferrara2016rise,cresci2020decade,crescietal2017paradigm}. As detection tools
incorporate particular behavioural signals (e.g.\ excessive URLs, heavy
duplication, lack of emojis or media), it is plausible that bot operators
adjust those signals to appear more human-like. Our longitudinal analysis
shows that such adjustments are not merely anecdotal but are visible at scale
across thousands of promotional bots.

\subsection{Generational shifts and life-cycle strategies}

The generational and age-class analyses reveal more nuanced patterns of
adaptation. Second-generation bots combine high volume, heavy link sharing,
and extensive use of duplicated text and hashtags, consistent with naive but
highly aggressive promotional strategies. Third-generation bots, by contrast,
exhibit more sophisticated expressive behaviour---richer use of emojis and
media, more nuanced sentiment---and somewhat reduced use of obvious signals
such as extreme duplication and multi-language broadcasts. This echoes prior
observations that newer bots often look increasingly similar to genuine users
\citep{pozzana2020measuring,puertas2019bots}.

Differences between short-lived and long-lived bots suggest distinct
life-cycle strategies. Short-lived bots appear to be created for targeted,
time-limited campaigns, operating at very high intensity and then disappearing
or being suspended. Long-lived bots, in contrast, adopt more moderate and
adaptive behaviours, potentially allowing them to avoid detection and survive
for many years. Similar life-cycle patterns have been noted for spam and other
malicious accounts \citep{lee2011sevenmonths}.

\subsection{Implications for bot detection}

Our findings have several implications for bot-detection research and practice:

\begin{enumerate}
  \item \textbf{Temporal drift in feature distributions.} Detection models
  trained on historical data may rely on feature distributions that no longer
  hold. If bots steadily increase their use of media, emojis, or certain
  sentiment patterns, classifiers that treat extreme values as suspicious may
  degrade in performance or misclassify humans.
  \item \textbf{Need for time-aware models.} Incorporating temporal
  information---either through explicit time-series models
  \citep{pozzana2020measuring} or through periodic retraining and concept-drift
  adaptation---is important for robust bot detection. Models should account
  for trends in both bot and human behaviour.
  \item \textbf{Cohort-specific behaviours.} Behavioural signatures differ
  across generations and age classes. Classifiers might benefit from assigning
  accounts to cohorts and applying cohort-specific models, or from including
  activation time and lifespan as features in their own right.
  \item \textbf{Beyond static snapshots.} Many current approaches operate on
  static snapshots of activity (e.g.\ a user's last 200 tweets). Our results
  suggest that detection systems should consider the evolution of behaviour
  over time, not just cross-sectional differences at a single point.
\end{enumerate}

This study has presented a longitudinal analysis of the behaviour of
promotional Twitter bots over a 12-year period. Using ten content-based
meta-features and standard time-series tests, we have shown that these bots
do \emph{not} exhibit stationary behaviour: their activity volumes, content
richness, and expressive strategies change systematically over time. Generational
and age-class analyses further reveal distinct patterns between older and newer
bots and between short-lived and long-lived accounts.

The findings reinforce the view that social bots are dynamic adversaries and
that bot-detection systems must account for temporal drift in behavioural
features. 

To explore this, we undertook a  second study to better understand the relationship between characteristics of bot behaviour and so build a better detection approach.

\section{Feature relationships in bots to aid detection}

\subsection{Feature engineering for bot detection}

Bot detection systems leverage a range of feature types, often grouped into profile, content, temporal, and network categories. 
Profile features include account age, number of followers and friends, and profile description properties. 
Content features capture aspects of the text itself, such as n-grams, part-of-speech patterns, sentiment, and use of URLs and hashtags. 
Temporal features consider posting rates, inter-tweet times, and time-of-day distributions, while network features consider retweet, mention, and follower graphs.  

Prior work has proposed specialised feature sets tailored to particular bot families or campaigns, such as content duplication patterns, coordinated retweet bursts, or client application metadata. 
De~Nicola et al.\ compare four feature sets - including Botometer scores, profile/timeline features, and Twitter client information - for detecting new bots and show that no single set dominates \citep{nicola_efficacy_2021}. 
Rovito et al.\ use genetic programming to evolve interpretable classification rules based on feature subsets, demonstrating that evolutionary computation can discover compact feature combinations that perform competitively on modern datasets like TwiBot-20 \citep{rovito_evolutionary_2022}. 
Gilmary et al.\ represent users' posting histories as symbolic sequences and apply relative entropy metrics to identify automated behaviour with promising performance and minimal training overhead \citep{gilmary_dna-influenced_2022}.

Despite this breadth, most feature engineering work still evaluates features on static datasets. 
Our study complements these efforts by focusing on the \emph{temporal stability} of feature relationships: rather than proposing a new feature set, we ask how the same set behaves when applied to multiple bot generations.

\subsection{Coordinated and group-level behaviour}

Beyond individual accounts, recent research examines coordinated campaigns and group-level structures. 
Pacheco et al.\ present a framework for uncovering coordinated networks on social media by building similarity graphs from shared content and interaction patterns \citep{pacheco_uncovering_2021}. 
Their approach reveals tightly knit groups of accounts---often bots---that act in concert during political events or information operations. 
Other work explores how behavioural signatures differ between bots and humans at the group level, including clustering of posting behaviour, sentiment, and topical focus \citep{mouronte-lopez_patterns_2024}. 

Our analysis is not a coordination detector, but it sits in a similar methodological space: we focus on how combinations of actions (tweeting vs.\ retweeting), URLs, hashtags, sentiment, emojis, and media co-occur within and across bot generations. 
Changes in these combinations over time can be interpreted as evidence of strategic adaptation, akin to the evolving tactics observed in coordinated networks.

\subsection{Study aims}

The study is driven by the following research question:

\medskip
\noindent\textbf{RQ1:} \emph{Is there evidence of dynamic changes in the relationships between behavioural features of promotional Twitter bots across successive generations?}
\medskip

Here, ``relationships'' refer to statistical dependencies and correlations between binary or categorical behavioural features derived from tweets, such as whether bots that use many hashtags are also more likely to attach media, or whether sentiment is associated with URL usage. 
By examining these relationships across three generations of bots deployed for the same or similar campaigns, we can infer whether bot developers have altered how features are combined.

The key features of our study were as follows:
\begin{itemize}
  \item We develop an interpretable behavioural feature set for promotional Twitter bots, covering posting actions, topical similarity, URLs, hashtags, sentiment, emojis, and media (Section~\ref{sec:data-features}).
  \item We present a three-step framework (Figure~\ref{fig:framework}) combining dependency analysis, correlation tracking, and evolution pattern synthesis to study how relationships among features change across three bot generations (Section~\ref{sec:method}).
  \item We provide an empirical characterisation of feature relationships across generations (Section~\ref{sec:results}), showing widespread dependency (almost all feature pairs are dependent), systematic shifts in correlation strength and polarity, and distinctive evolution patterns for key feature combinations.
  \item We connect these findings to recent work on evolving bots and model robustness \citep{nicola_efficacy_2021,rovito_evolutionary_2022,gilmary_dna-influenced_2022,mouronte-lopez_patterns_2024}, and discuss implications for the stability of behavioural features in practical detection systems (Section~\ref{sec:discussion}).
\end{itemize}

\section{Data and Feature Engineering}
\label{sec:data-features}

\subsection{Dataset and bot generations}

The analysis uses a labelled dataset of promotional Twitter bots collected from multiple campaigns. 
Bots are grouped into three \emph{generations} (G1, G2, G3), corresponding to distinct deployment periods and design iterations. 
Each generation contains accounts that promote similar targets but were created and operated at different times. 
The underlying thesis chapter (from which this paper is derived) provides full details of data collection and labelling; here we focus on the feature-engineering steps relevant to behaviour dynamics.

Tweets from bot accounts are pre-processed to remove obvious noise (e.g., malformed messages, deleted content), and then aggregated at the level of individual tweets with associated metadata (hashtags, URLs, emojis, media). 
Language and timestamp features are excluded from the behaviour-dynamics analysis to avoid confounding factors and to focus on structural aspects of posting behaviour.

\subsection{Behavioural feature set}

We design an interpretable feature set capturing key aspects of promotional bot behaviour. 
Seven high-level factors are considered (posting action, text similarity, URL usage, hashtags, sentiment, emojis, media), which are further decomposed into 18 binary features. 
Table~\ref{tab:features} summarises the feature definitions.

\begin{table}[htb]
\centering
\caption{Behavioural features used to characterise promotional Twitter bots. 
Each feature is binary; multi-valued factors are decomposed into several binary indicators.}
\label{tab:features}
\begin{tabular}{p{3cm}p{9cm}}
\toprule
\textbf{Category} & \textbf{Features in this study} \\
\midrule
Posting action &
\textbf{F1: Tweet only} (original tweets, no retweets/replies); 
\textbf{F2: Retweet only} (retweets only); 
\textbf{F3: Tweet and retweet} (both original tweets and retweets/replies). \\[0.4em]
Text similarity &
\textbf{F4: Single-topic tweets} (high lexical similarity, near-duplicate promotional content); 
\textbf{F5: Mixed-topic tweets} (distinct topics or templates); 
\textbf{F6: Topic-infrequent tweets} (idiosyncratic or low-frequency topics). \\[0.4em]
URL usage &
\textbf{F7: URL present} (tweet includes at least one URL); 
\textbf{F8: No URL} (no URLs). \\[0.4em]
Hashtags &
\textbf{F9: Single hashtag} (exactly one hashtag); 
\textbf{F10: Multiple hashtags} (two or more); 
\textbf{F11: No hashtag}. \\[0.4em]
Sentiment polarity &
\textbf{F12: Positive sentiment}; 
\textbf{F13: Negative sentiment}; 
\textbf{F14: Neutral or unclear sentiment}. \\[0.4em]
Emojis &
\textbf{F15: Emoji present}; 
\textbf{F16: No emoji}. \\[0.4em]
Media &
\textbf{F17: Media present} (image, GIF, or video attached); 
\textbf{F18: No media}. \\[0.4em]
Generation &
Each tweet is also associated with its bot generation (G1, G2, G3), which is used to segment the data for cross-generation comparisons. \\
\bottomrule
\end{tabular}
\end{table}

The feature set was designed to be:
\begin{itemize}
  \item \emph{Behavioural}: focusing on what bots do (actions and content structure), not on user IDs or network structure.
  \item \emph{Interpretable}: allowing qualitative interpretation of patterns and evolution (e.g., shifts from single to multiple hashtags).
  \item \emph{Comparable}: applicable across all three generations without needing re-engineering as platforms or APIs change.
\end{itemize}

\section{Methodology}
\label{sec:method}

\begin{figure}[htb!]
\centering
\includegraphics[width=0.6\textwidth]{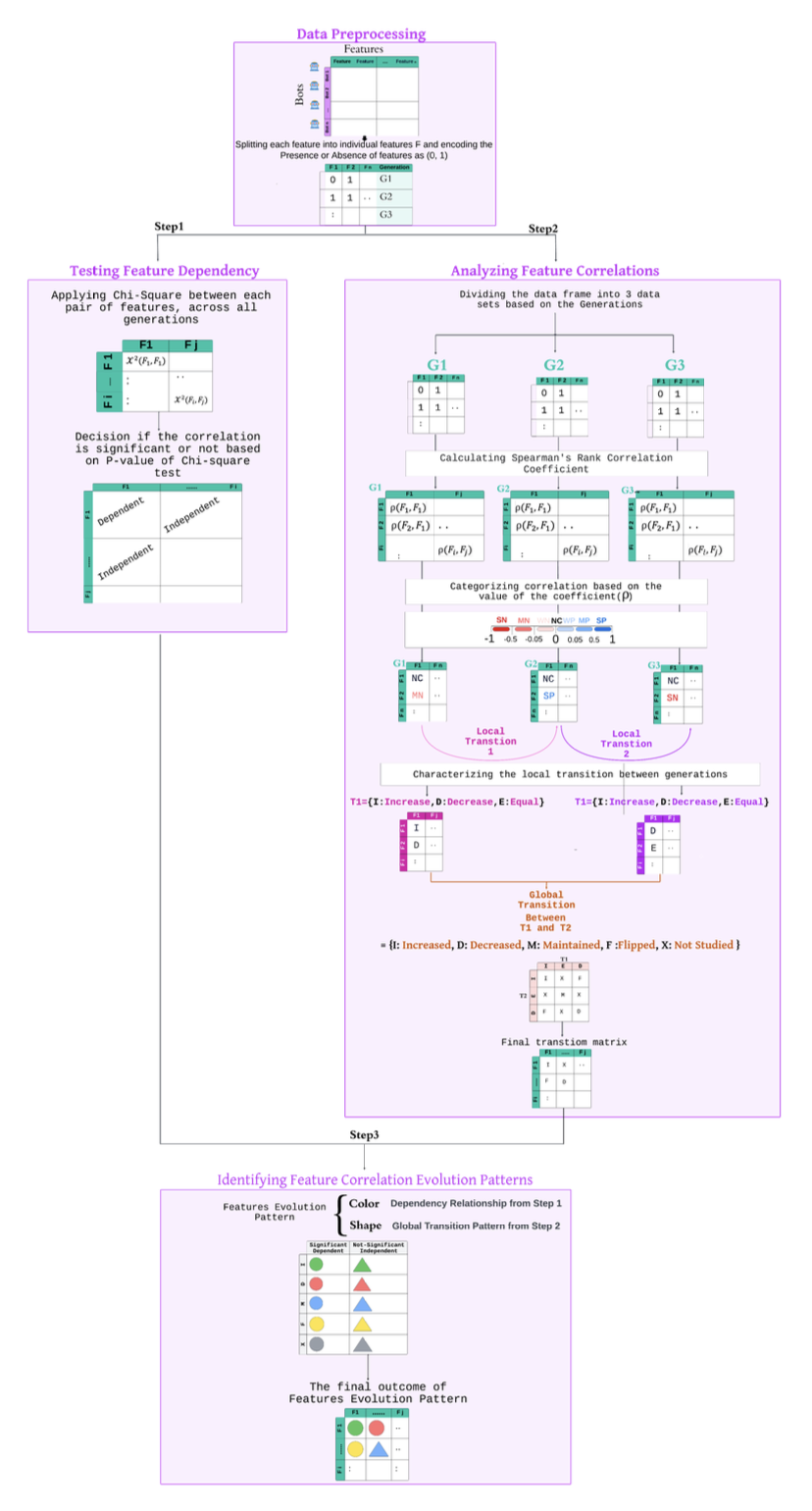}
\caption{Analytical framework for characterising behaviour dynamics in promotional Twitter bots. 
Step~1 computes pairwise dependencies between features.
Step~2 quantifies correlations per generation and tracks local/global transitions.
Step~3 aggregates these into qualitative evolution patterns.}
\label{fig:framework}
\end{figure}

To answer RQ1, we employ a three-step framework (Figure~\ref{fig:framework}):

\begin{enumerate}
  \item Evaluate pairwise \emph{feature dependencies} across the entire dataset using $\chi^2$ tests.
  \item For each generation, compute pairwise correlations between features and categorise them into discrete strength/polarity classes; then track \emph{local} (between successive generations) and \emph{global} (across all generations) transitions.
  \item Synthesize these transitions into \emph{behaviour evolution patterns} that describe how feature relationships change over time.
\end{enumerate}

\subsection{Step 1: Feature dependency analysis}

Given the binary features in Table~\ref{tab:features}, we consider all unordered feature pairs $(F_i,F_j)$ with $i \neq j$. 
With 18 features, this yields $18 \choose 2$ $= 153$ distinct pairs. 
For each pair, we construct a $2\times 2$ contingency table over all tweets from all generations and compute the $\chi^2$ statistic and $p$-value. 
We then classify pairs as \emph{dependent} or \emph{independent} using a significance threshold of $\alpha=0.05$ (with Bonferroni correction applied in the original thesis chapter to account for multiple testing).

\begin{figure}[htb!]
\centering
\includegraphics[width=0.8\textwidth]{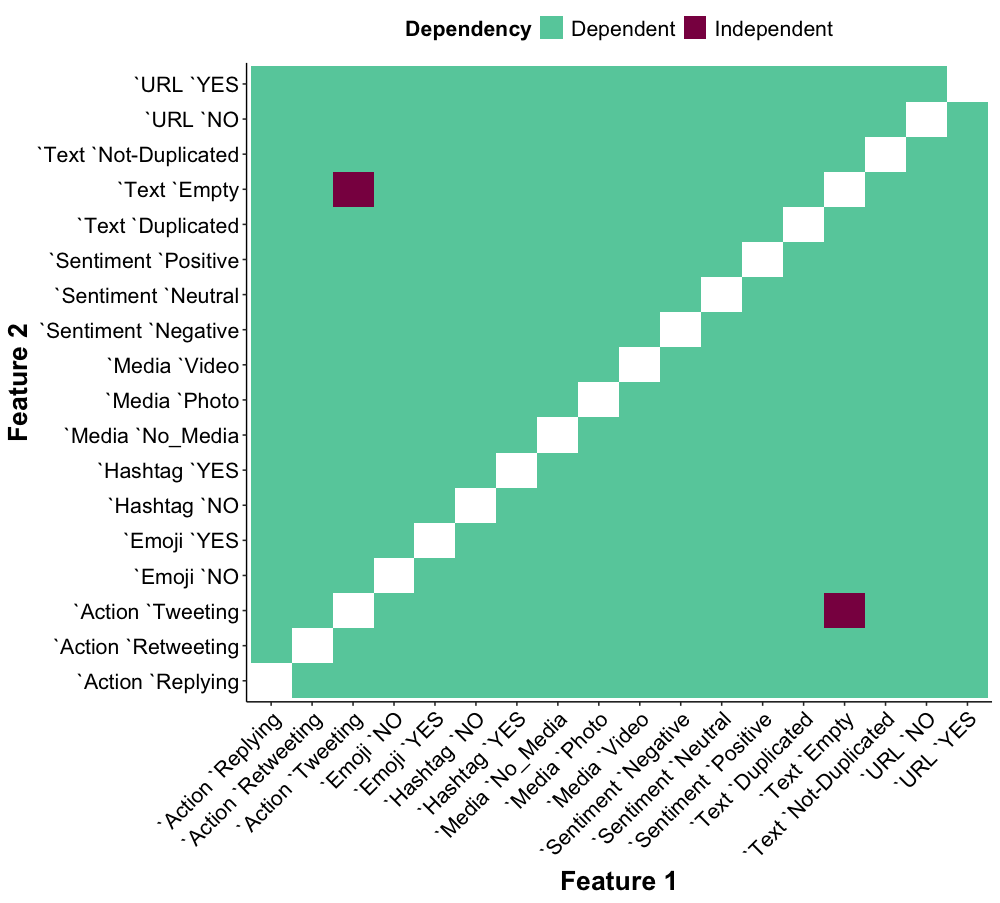}
\caption{Schematic $\chi^2$ dependency matrix across all feature pairs. 
Almost all pairs of behavioural features show statistically significant dependence, indicating strong interrelations between posting actions, hashtags, URLs, sentiment, emojis, and media.}
\label{fig:chi2matrix}
\end{figure}

The resulting dependency matrix can be visualised as a heat map where each cell corresponds to a feature pair and is shaded according to significance and effect size (Cram\'er's V). 
Figure~\ref{fig:chi2matrix} conceptually illustrates this matrix: darker cells indicate stronger dependence.

\subsection{Step 2: Correlation categories and transitions}

While dependency analysis tells us whether two features are associated, it does not distinguish positive from negative associations or indicate how strong they are in each generation. 
We therefore compute pairwise correlations within each generation.

For each generation $g \in \{1,2,3\}$ and each pair $(F_i,F_j)$, we estimate the Spearman rank correlation coefficient $\rho_g(F_i,F_j)$ over tweets belonging to bots in that generation.\footnote{Because the features are binary, Spearman and Pearson correlations are equivalent up to scaling; Spearman is retained for consistency with the thesis chapter.}
To simplify comparisons, we discretise correlation values into categories:

\begin{itemize}
  \item \textbf{SP} (Strong Positive): $\rho \geq \tau_{\text{SP}}$
  \item \textbf{MP} (Moderate Positive): $\tau_{\text{MP}} \leq \rho < \tau_{\text{SP}}$
  \item \textbf{WP} (Weak Positive): $0 < \rho < \tau_{\text{MP}}$
  \item \textbf{NP} (No Correlation): $\rho \approx 0$ (within a small band around zero)
  \item \textbf{WN} (Weak Negative): $-\tau_{\text{MP}} < \rho < 0$
  \item \textbf{MN} (Moderate Negative): $-\tau_{\text{SP}} < \rho \leq -\tau_{\text{MP}}$
  \item \textbf{SN} (Strong Negative): $\rho \leq -\tau_{\text{SP}}$
\end{itemize}

with thresholds (e.g.\ $\tau_{\text{SP}} = 0.7$, $\tau_{\text{MP}} = 0.3$) chosen to match conventional interpretations used in the thesis chapter. 

For each generation $g$, we thus obtain a categorical correlation matrix $C_g$ where entry $C_g(i,j)$ is one of $\{\text{SP},\text{MP},\text{WP},\text{NP},\text{WN},\text{MN},\text{SN}\}$. 
Aggregating over all 153 pairs yields a distribution of categories per generation (e.g.\ how many pairs are moderately positive in G1 vs.\ G2).

We then define:
\begin{itemize}
  \item \textbf{Local transitions}: for each consecutive pair of generations $(g,g+1)$ and each feature pair $(i,j)$, we consider the ordered pair $\big(C_g(i,j), C_{g+1}(i,j)\big)$, describing how that relationship changes between those two generations. 
  \item \textbf{Global transitions}: for each feature pair $(i,j)$, we consider the triplet $\big(C_1(i,j), C_2(i,j), C_3(i,j)\big)$, capturing its entire trajectory across generations.
\end{itemize}

Local transitions can be summarised in a matrix where rows and columns correspond to correlation categories in the earlier and later generation respectively (e.g.\ WP$\rightarrow$MP, MP$\rightarrow$SP, etc.), and cells contain counts of how many feature pairs follow each pattern. 

\begin{figure}
    \centering
    \includegraphics[width=0.9\linewidth]{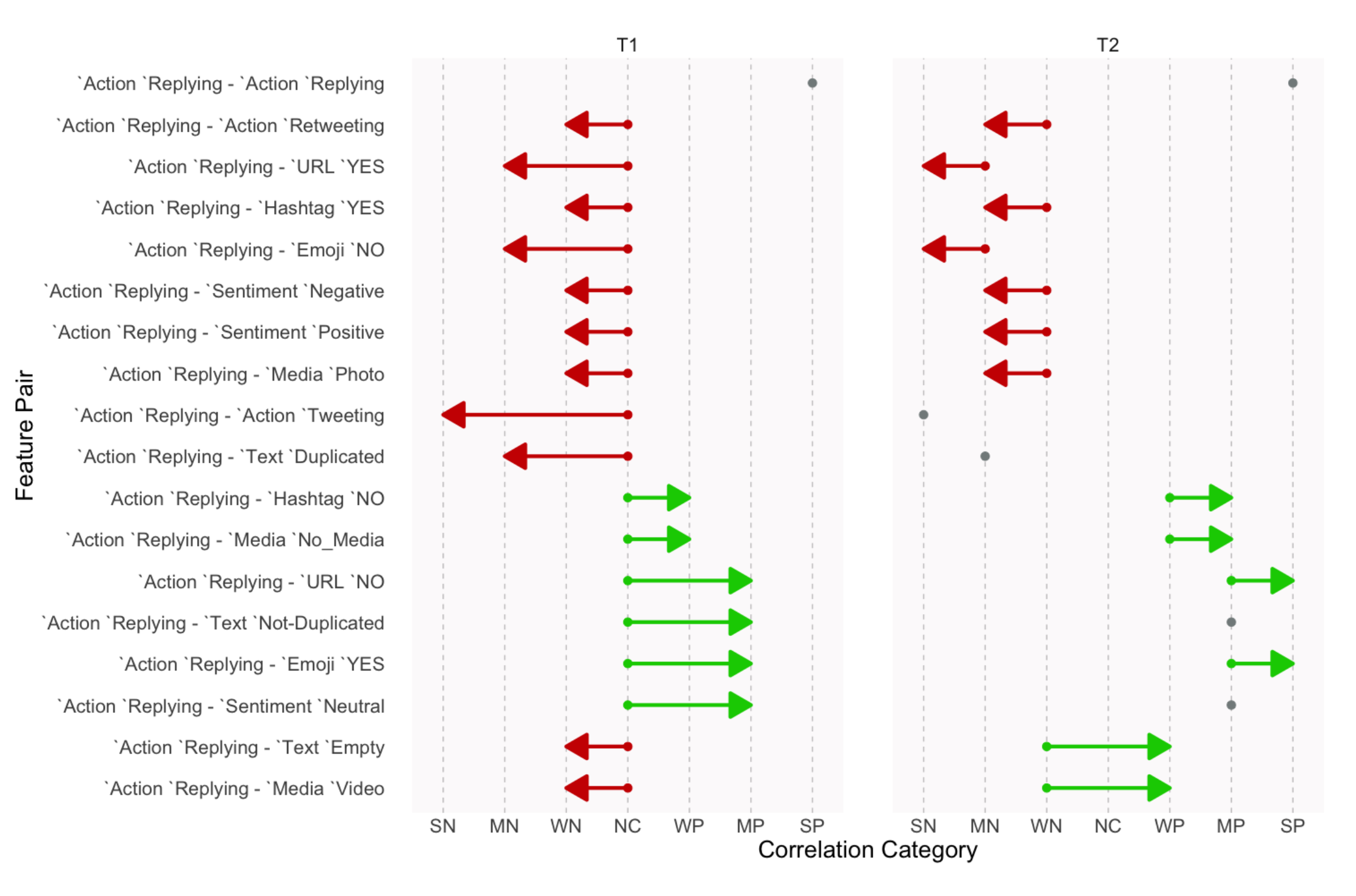}
    \caption{Local Transitions T1 and T2 express changes in correlations between
bots’ features over three generations, for 18 key transitions}
    \label{fig:transitions}
\end{figure}

Figure \ref{fig:transitions} is an arrow plot that visually represents local transitions: T1 (between
the first and second generations) and T2 (between the second and third gener-
ations). The plot illustrates how the correlation between each pair of features
evolves across generations, categorized as follows: increased (green arrows), de-
creased (red arrows), or remained the same (dots). While the figure highlights the most significant \textbf{\textit{(is this the case? RB)}} 18 feature pairs, the remaining pairs are not presented owing to space limitations.
\subsection{Step 3: Behaviour evolution patterns}

The local and global transitions are finally synthesised into qualitative \emph{evolution patterns}. 
For example:
\begin{itemize}
  \item Pairs that remain in the same category across all generations (e.g.\ WP$\rightarrow$WP$\rightarrow$WP) are labelled \emph{stable}.
  \item Pairs that move from weaker positive to stronger positive (e.g.\ WP$\rightarrow$MP$\rightarrow$MP or MP$\rightarrow$SP) are labelled \emph{strengthening positive}.
  \item Pairs that flip from positive to negative (e.g.\ WP$\rightarrow$WN or MP$\rightarrow$MN) are labelled \emph{sign reversal}.
  \item Pairs with back-and-forth changes (e.g.\ WP$\rightarrow$MP$\rightarrow$WP) are labelled \emph{variable}.
\end{itemize}

These patterns are used in Section~\ref{sec:results} to interpret which feature combinations become more tightly coupled, which decouple, and which change sign, offering insight into how bot developers adjust their strategies.

\section{Results}
\label{sec:results}

\subsection{Feature dependencies}

The $\chi^2$ analysis reveals that almost all of the 153 feature pairs show statistically significant dependency across the combined dataset. 
Only a single pair fails to reach the corrected significance threshold. 
This implies that behavioural features do not operate independently: choices about posting actions, hashtags, URLs, sentiment, emojis, and media tend to co-occur in structured ways.

The strongest dependencies involve:
\begin{itemize}
  \item Posting action and hashtag usage: bots that retweet heavily exhibit different hashtag patterns compared to those that primarily post original tweets.
  \item URL usage and media: tweets with URLs often differ in media attachment patterns compared to those without URLs.
  \item Hashtags and sentiment: tweets with richer hashtag usage show distinct sentiment profiles relative to tweets without hashtags.
\end{itemize}

Figure~\ref{fig:chi2matrix} (from the thesis chapter) visually summarises these dependencies as a dense grid of significant associations.

\subsection{Correlation categories across generations}

When correlations are examined separately for each generation, we observe clear differences in the distribution of correlation categories. 
In the earliest generation (G1), most feature pairs fall into weak positive or weak negative categories, with very few strong correlations in either direction. 
By the second and third generations, the number of moderate and strong correlations increases, particularly for pairs involving hashtags, URLs, and media. 
At the same time, several relationships move from weakly positive or near-zero correlations in G1 to weakly or moderately negative correlations in later generations.

Conceptually, the distribution across categories can be described as follows:
\begin{itemize}
  \item \textbf{G1}: dominated by weak correlations (both positive and negative), many pairs close to zero, almost no strong positive relationships.
  \item \textbf{G2}: increased number of moderate positive and moderate negative correlations; a small number of pairs become strongly positive.
  \item \textbf{G3}: further increase in moderate and strong correlations (both positive and negative), indicating increasingly structured coordination between features.
\end{itemize}

These distributions are summarised in Table \ref{tab:correlation-categories} counting how many of the 153 pairs fall into each category for each generation.
\begin{table}[htb!]
\centering
\caption{Distribution of correlation categories across bot generations}
\label{tab:correlation-categories}
\begin{tabular}{lccc}
\hline
\textbf{Correlation Category} & \textbf{First Generation} & \textbf{Second Generation} & \textbf{Third Generation} \\
\hline
WP (Weak Positive)   & 34 & 28 & 13 \\
WN (Weak Negative)   & 34 & 33 & 12 \\
NC (No Correlation)  & 33 &  0 &  0 \\
MN (Moderate Negative) & 25 & 44 & 61 \\
MP (Moderate Positive) & 20 & 40 & 52 \\
SP (Strong Positive) &  0 &  0 &  4 \\
SN (Strong Negative) &  7 &  8 & 11 \\
\hline
\end{tabular}
\end{table}

\subsection{Local transitions between generations}

Local transitions capture how individual feature pairs change between G1 and G2, and between G2 and G3. 
The main observations are:

\begin{itemize}
  \item The most frequent pattern is \emph{stability within a weak or moderate category} (e.g.\ WP$\rightarrow$WP, MP$\rightarrow$MP), reflecting a core set of relationships that remain broadly similar across generations.
  \item Among non-stable transitions, there is a clear predominance of \emph{strengthening} trajectories, especially WP$\rightarrow$MP and MP$\rightarrow$SP for positive correlations, and WN$\rightarrow$MN for negative correlations. 
        This suggests that certain hedged relationships become more pronounced over time.
  \item \emph{Sign reversals} (e.g.\ WP$\rightarrow$WN, MP$\rightarrow$MN) occur but are less frequent; they often involve combinations of sentiment with hashtags or URLs, indicating that bots adjust how sentiment is packaged with promotional markers.
\end{itemize}

\subsection{Global transition patterns across generations}

Considering the full trajectories from G1 to G3, we can group feature pairs into broad evolution patterns:

\begin{itemize}
  \item \textbf{Stable weak relationships}: many pairs remain weakly positive or weakly negative throughout (e.g.\ WP$\rightarrow$WP$\rightarrow$WP, WN$\rightarrow$WN$\rightarrow$WN). 
        These may reflect background tendencies in promotional behaviour that are not heavily tuned by bot developers.
  \item \textbf{Strengthening positive relationships}: several pairs move from weak to moderate or strong positive correlations (e.g.\ WP$\rightarrow$MP$\rightarrow$MP, MP$\rightarrow$SP$\rightarrow$SP). 
        Examples include the coupling between multiple hashtags and media presence, suggesting that later generations systematically co-use these features.
  \item \textbf{Emerging negative relationships}: some pairs evolve from near-zero or weak positive correlations in G1 to weak or moderate negative correlations in later generations (e.g.\ NP$\rightarrow$WN$\rightarrow$MN). 
        These often involve sentiment and specific combinations of hashtags or URLs, indicating strategic diversification or division of labour among bot accounts.
  \item \textbf{Variable relationships}: a smaller set of pairs exhibits back-and-forth changes (e.g.\ WP$\rightarrow$MP$\rightarrow$WP), possibly reflecting experimentation or inconsistent strategies.
\end{itemize}

These patterns are summarised as a global transition table (Table \ref{tab:global-transitions}, which counts how many pairs belong to each evolution pattern. 
\begin{table}[htb!]
\centering
\caption{Summary of the correlation count of global transitions}
\label{tab:global-transitions}
\begin{tabular}{cccc}
\toprule
T1 & T2 & Count & Global transition \\
\midrule
E & E & 29 & Equal \\
D & D & 19 & Decreased \\
I & I & 15 & Increased \\
D & I & 12 & Flipped \\
I & D & 11 & Flipped \\
D & E & 26 & Not studied \\
I & E & 22 & Not studied \\
E & I & 11 & Not studied \\
E & D & 8  & Not studied \\
\bottomrule
\end{tabular}
\end{table}

\begin{figure}[htb!]
\centering
\includegraphics[width=0.9\textwidth]{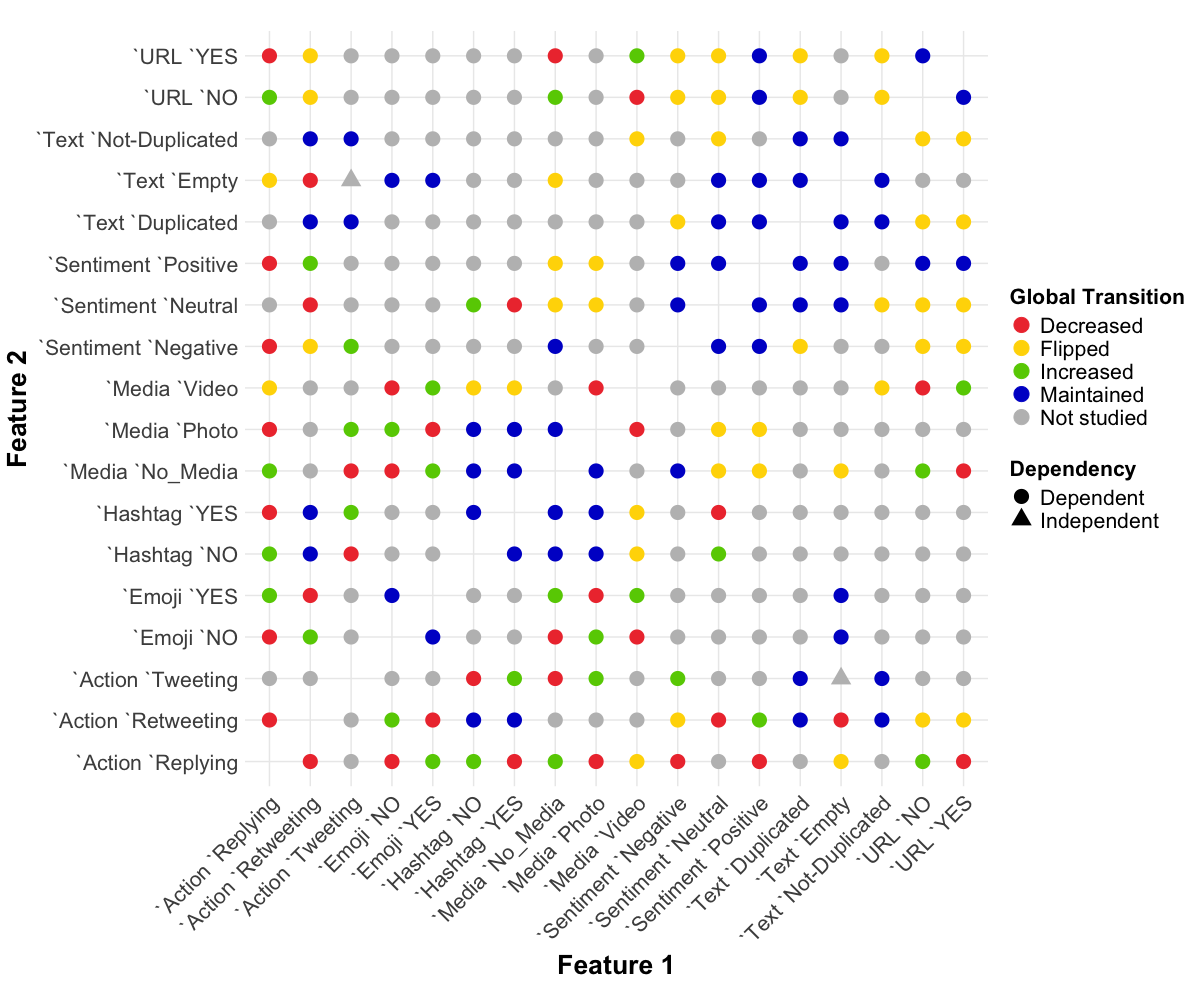}
\caption{Schematic illustration of global evolution patterns for pairwise feature relationships. 
Each cell corresponds to a feature pair; colour encodes correlation polarity (positive vs.\ negative) and shape encodes qualitative trend (stable, strengthening, sign reversal, variable).}
\label{fig:evol-patterns}
\end{figure}

A companion figure (Figure~\ref{fig:evol-patterns}) uses colours and shapes to encode these patterns for all 153 pairs.

\subsection{Behaviour evolution insights}

Combining the dependency and correlation analyses yields several qualitative insights into how promotional bot behaviour evolves:

\paragraph{From volume to structure.}  
Early-generation bots rely heavily on volume and repetition: many relationships among hashtags, URLs, and media are only weakly correlated, reflecting simple templates repeatedly broadcast. 
Later generations exhibit stronger and more differentiated correlations, suggesting more carefully structured posts that combine multiple cues (e.g.\ specific hashtags plus images) in consistent ways.

\paragraph{Refined hashtag strategies.}  
Hashtag usage becomes more systematically tied to other features. 
For example, the association between multiple hashtags and media presence strengthens, while the relationship between ``no hashtags'' and certain sentiment categories becomes more clearly negative. 
This indicates that bots increasingly use hashtags in a targeted manner rather than indiscriminately.

\paragraph{Sentiment packaging.}  
While promotional content is generally positive, the way sentiment combines with URLs and hashtags changes over time. 
Some relationships between positive sentiment and promotional markers weaken, while others between neutral sentiment and specific hashtag patterns strengthen. 
This may reflect attempts to avoid overly enthusiastic language that might be flagged as spam, while still maintaining marketing objectives.

\paragraph{Experimentation with emojis and media.}  
Emojis and media attachments show mixed evolution patterns. 
Some combinations (e.g.\ media with multiple hashtags) strengthen, while others (emoji usage with certain text-structure features) remain weak or variable. 
This suggests that developers experiment with expressive elements, but only some of these experiments lead to stable strategies.

\section{Discussion}
\label{sec:discussion}

\subsection{Evidence for evolving bot behaviour}

The main research question asked whether there is evidence of dynamic changes in relationships between behavioural features of promotional Twitter bots across generations. 
Taken together, the results support an affirmative answer:

\begin{itemize}
  \item Almost all feature pairs are dependent overall, indicating a complex web of behavioural couplings.
  \item The distribution of correlation strengths and signs changes across generations, with more moderate and strong correlations in later generations.
  \item Local and global transition patterns show a bias towards strengthening relationships and, to a lesser extent, sign reversals, particularly for combinations involving hashtags, URLs, media, and sentiment.
\end{itemize}

These patterns are consistent with the idea that bot developers refine their use of behavioural cues over time, moving from relatively ad-hoc templates towards more structured and coordinated strategies that may be harder to detect with static feature sets.

\subsection{Implications for bot detection and feature robustness}

Our findings resonate with recent concerns about the robustness of detection features \citep{nicola_efficacy_2021}. 
If relationships between features evolve, detection models trained on older data may misestimate the importance of certain feature combinations and underperform on newer bots. 
For example, if early models rely on the strong co-occurrence of URLs and certain hashtag patterns, but later bots decouple these cues, the models may become brittle.

At the same time, the presence of stable relationships suggests that some behavioural features have enduring discriminative power. 
De~Nicola et al.\ argue that inexpensive profile and timeline features can still be effective on new bots \citep{nicola_efficacy_2021}; our results complement this by highlighting which combinations of content-level features appear relatively stable and which evolve. 
Recent proposals that incorporate richer sequence-based representations \citep{gilmary_dna-influenced_2022} or evolve classifiers using genetic programming \citep{rovito_evolutionary_2022} could benefit from explicitly monitoring feature relationships over time and adapting models accordingly.

\subsection{Relation to group-level and domain-specific studies}

Studies of coordinated networks and topic-specific bots often reveal structured behaviour within campaigns \citep{pacheco_uncovering_2021,mouronte-lopez_patterns_2024}. 
Our analysis, focused on promotional campaigns, shows a similar tendency: later bot generations exhibit more coordinated combinations of hashtags and media, and more deliberate sentiment packaging. 
This suggests that the evolution observed here may be part of a broader pattern in which bots become both more sophisticated and more specialised.

\subsection{Limitations}

Several limitations should be noted:

\begin{itemize}
  \item \textbf{Feature selection.}  
        The feature set is intentionally narrow and interpretable, focusing on a subset of content-level behaviours. 
        Other important signals (e.g.\ temporal rhythms, follower networks, client applications) are not included in the dynamics analysis.
  \item \textbf{Binary encoding.}  
        Discretising features into binary indicators simplifies interpretation but can obscure nuanced variation (e.g.\ a wide range of posting rates collapsed into a high/low threshold).
  \item \textbf{Correlation categorisation.}  
        The thresholds used to define correlation categories are somewhat arbitrary, and alternative thresholding schemes could yield slightly different transition counts. 
        However, the qualitative patterns (movement from weaker to stronger categories, emergence of negative relationships) are robust.
  \item \textbf{Dataset scope.}  
        The dataset focuses on specific promotional campaigns and three generations of bots. 
        Results may not generalise to other domains (e.g.\ political bots, conversational bots) without further validation.
\end{itemize}

\subsection{Future work}

Future research could extend this analysis in several directions:

\begin{itemize}
  \item Incorporate temporal and network features into the evolution framework to capture not only what bots post but when and in which interaction structures.
  \item Apply the framework to mixed human--bot populations to compare how feature relationships evolve for bots versus humans over the same period.
  \item Integrate behaviour-dynamics monitoring into online detection systems, allowing models to adjust feature weights or architectures when significant changes in feature relationships are detected.
  \item Study post-2020 bot generations, especially in contexts where large language models are used to generate content, to see whether behaviour dynamics become more or less distinctive.
\end{itemize}

\section{Summary}

This paper has examined the evolution of behavioural feature relationships in promotional Twitter bots across three generations. 
Using a simple, interpretable feature set and a three-step analytical framework, we showed that almost all feature pairs exhibit dependency, and that the strength and polarity of these dependencies change in systematic ways over time. 
Later generations display more structured combinations of hashtags, URLs, media, and sentiment, consistent with progressively refined strategies.

These findings contribute to a growing literature on evolving bots and detection robustness. 
They suggest that monitoring how feature relationships change can provide early warning signals of new bot strategies and can inform the design of more resilient detection systems. 
As automation and generative models continue to advance, such behaviour-dynamics perspectives will be increasingly important for understanding and countering malicious and manipulative activity online.

\subsection{Limitations}

Several limitations should be acknowledged. First, the dataset focuses on promotional bots drawn from established research
corpora. These bots have been detected and labelled in earlier work; their
behaviour may differ systematically from that of undetected or more
sophisticated bots that remain hidden. Our findings therefore speak most
directly to the evolution of \emph{known} promotional bots.

Second, data collection is limited to the most recent 3{,}200 tweets per
account, which may under-represent the earliest behaviour of the oldest bots.
This constraint may attenuate early-year trends.

Third, we aggregate behaviour at yearly resolution, which simplifies temporal
patterns and may obscure seasonal dynamics, bursty campaign phases, or
short-lived anomalies. More fine-grained temporal modelling (e.g.\ monthly or
event-centred) could reveal additional structure.

Fourth, our stationarity and trend analyses focus on linear trends and do not
explicitly model potential nonlinearities or structural breaks. Bot behaviour
may shift abruptly in response to major platform changes, policy updates, or
external events (such as the COVID-19 pandemic), which are only indirectly
captured in our framework.

Finally, Twitter as a platform has undergone substantial changes over the
study period, including modifications to its interface, algorithms, and
enforcement practices. These changes likely influence both bot and human
behaviour, and disentangling platform effects from bot adaptation remains an
open challenge.  These changes have continued as it has been taken private, but there is little reason to believe tha the results of this study are invalidated.

\section{Conclusion and Future Work}

tbc

\bigskip
\noindent\textbf{Acknowledgements.}
Ohoud Alzahrani undertook the experiments and did the data analysis.  Bob Hendley and Ohoud Alzahrani devised the experiments, with input from Russell Beale, who wrote the paper based on the thesis of Ohoud Alzahrani. Generative AI was used to extract information from the thesis pdf, such as images, and assist with revising the literature review and identifying other relevant work.  The authors are responsible for the text.

Data availability: The dataset for these experiments may be available by contacting Ohoud Alzharani by email.

\bigskip
\bibliographystyle{abbrvnat}
\bibliography{refs}

\end{document}